# Multidimensional User Data Model for Web Personalization

Nithin K. Anil
M.A. College of Engineering
Kothamangalam
Kerala, India

Sharath Basil Kurian
School of Engineering
CUSAT, Kochi
Kerala, India

Aby Abahai T.
M.A. College of Engineering
Kothamangalam
Kerala, India

Surekha Mariam Varghese
M.A. College of Engineering
Kothamangalam
Kerala, India

## ABSTRACT

Personalization is being applied to great extend in many systems. This paper presents a multi-dimensional user data model and its application in web search. Online and Offline activities of the user are tracked for creating the user model. The main phases are identification of relevant documents and the representation of relevance and similarity of the documents. The concepts Keywords, Topics, URLs and clusters are used in the implementation. The algorithms for profiling, grading and clustering the concepts in the user model and algorithm for determining the personalized search results by re-ranking the results in a search bank are presented in this paper. Simple experiments for evaluation of the model and their results are described.

## 1. INTRODUCTION

New websites are being launched everyday flooding the World Wide Web with similar contents. In 1995, there were only about ten thousand websites, according to Gray's statistics published at MIT site[1] and 16 millions of web users [2]. The count simply sky rocketed to around six hundred thousand in just two years. The World Wide Web crossed the mark of 1 million web sites in 2009 and 47 million websites were added in the same year. There 1.73 billion internet users today of which 1.4 billion are mail users with 247 billion mails are being sent every day. More than 4 billion photos are hosted by Flickr every month and 2.5 billion photos being uploaded to Facebook [3]. YouTube currently serves more than 4 billion video views every day[4]. All these information obtained from different surveys done by people over the internet shows that the amount of data that is being pumped into the internet day by day is astonishingly huge.

From this ever growing heap of data it is very difficult to hunt for a particular piece of information that is relevant to a particular user. For data retrieval, the user submits a search query to the search engine and manually picks the relevant links from list provided by the search engine. Usually search results are not tailored to the need of the particular user, but ordered based on many other factors that may not be relevant to the particular user. As a result, the user will have to browse through many pages to locate the relevant contents, even if it is present in the search results. Much research is going on to reduce the burden of the user by refining the search results according to the user needs. These systems are however not very efficient as they make a user data model based on the information obtained from how the users use their system. Currently personalization is used in many systems to a great extent. But the data model is separate and divided for each system. So Facebook profile of the user will be concentrating on the friendship details, Linked-in profile will be ba datased on the professional interests and so on. Mobasher et. al. has presented a personalization model integrating user transactions and page views [5].Our aim is to build a complete integrated and united profile portraying the diverse interests of the user which can be used in all variety of applications.

Web Personalization [6][7][8] enables customization or providing prioritized delivery of content based on the explicit or implicit interests of the individual user. Priority for a particular content or web page is determined by the details provided explicitly by the user or from the user's implicit/derived behavior and preferences such as links clicked or pages viewed.

Different implementations of web personalization are available now [9][10][11][12][13]. Personalized recommender systems have been greatly used for better performance of ecommerce systems. Collaborative filtering has been known to be one of the most successful recommendation methods. Collaborative filtering systems [14][15][16] collect preferences and interests of many users to make automatic predictions for similar users. Users will be asked to rate objects or mark their preferences so as to provide most suitable suggestions. This is based on the assumption that users with similar behavior have analogous interests [17]. Content-based filtering systems are solely based on individual users' preferences. The system tracks each user's behavior and recommends items to them that are similar to items the user liked in the past. In rule-based filtering [18] the users are asked to answer a sequence of questions derived from a decision tree. For more accurate predictions, a combination of the methods also may be attempted.

In this work we focus on data usage mining of the user with a view to make the Web experience of the user personalized to the user's taste. The experience can be something as casual as browsing the Web or something like trading stocks or purchasing something from an e-store. The proposed system uses a multidimensional user data model for web personalization. In addition to the explicit preferences specified by the user, offline and online activities of the user are considered for computing priority of the content. Activities of the user that are performed online such as during web browsing are termed as online activities. Data collected from online activities include the users browsing history, keystrokes and the click pattern followed during navigation of different pages. Location path, properties and types of files/applications frequently accessed by the user etc. are examples of offline activity data. Features derived from the user are stored in the central server for ensuring privacy and protect from misuse. No third party person, application or website can access the model without permission from the particular user. The user data models are further analyzed to identify the trending topics for a particular locality. This along with other information inferred collectively from a group of users will be kept open for all to view or use.

## 2. BACKGROUND

Web Directories, also referred to as Internet Directories or Knowledge Bases, are a popular means of organizing information resources on the web. A web directory is a repository of web pages that are organized in a hierarchical structure, usually like a tree or a directed acyclic graph (DAG). Each web page cataloged in a web directory is annotated with a short description. There is a hierarchical ordering for the pages. Each page or node is a special type of its parent node and generalized type of its children. Within each hierarchy, the non-leaf nodes specify a particular concept and the leaf nodes specify the sequence of web pages that are linked together on that concept. Every concept node may have any number of child nodes representing its sub-concepts. In DAG-structured web directories, a concept node may have more than one parent node. Users can locate information in a web directory by browsing through the concept hierarchy, identifying the relevant concepts and by examining the pages listed under the relevant concepts.

Google's Web Search indexes over eight billion pages. Instead of indexing the entirety of billions of pages, the directory describes sites instead, indexing about 1.5 million URLs [19]. Google Search Result Position Tool's are enhanced by the Google directory based on dmoz an open directory project. Full text of all the pages are indexed and stored in the index database of Google. The index structure is maintained as a sorted list of keywords. For each keyword, the index entries point to the corresponding keyword references in the document [20]. The links to the documents and alphabetically sorted list of keywords enables faster tracking of documents. Considering the huge size of the index structure, only important keywords are indexed by the Google Indexing service.

Many popular commercial search engines like Google, Bing and Yahoo have developed freely available APIs for accessing their index. Google has started offering free access to its index structure from 2002. Followed by Google, Yahoo and Microsoft also opened their index structure for public. This move has fired innovative research in the field of web search and now there is ample scope for refinements in the field.

## 3. MULTIDIMENSIONAL USER DATA MODEL

The purpose of user data modelling is to identify, index and prioritize the keywords that are relevant or important to a particular user. The browser history files are utilized for the indexing. The pages corresponding to different websites and stored documents are parsed to obtain the keywords in it. The searched keywords are also indexed. The priorities are given as per the importance of the keyword to the user. For example, searched keywords are given more priority than normal keywords. All the keywords obtained in this process are assigned relative priorities by comparing them with each other.

The multidimensional user data model was updated continuously by analyzing the behavior of the user in the system. The created model stores an approximation of user's interests. User models can be used to personalize systems to tailor generic content to the particular needs of a user. User models are updated automatically by tracking user's click stream, websites visited, documents and files in the system etc. The user data model is used to re-rank the list of objects issued to the user according to the user's implicit interests.

As shown in Figure 1, the system architecture has three main modules: crawler, analyzer, and ranker. When the system starts to run, it follows these steps. User browses the pages of his interest. As he visits or clicks on a particular page, the priority of the keywords in the particular page increases. The user can also dynamically set the priority of any page/keyword available on the Web. The crawler collects all keywords to keep the local collection fresh. A background process TRACK tracks visited pages and collects the keywords in it. All new pages are fetched and parsed for creating the keyword database. Each page will be describing some ideas or topics. These topics or ideas can be determined from the extracted keywords. Most relevant keyword is taken as the main topic. Once the keyword database gets updated, the analyzer runs the incremental keyword clustering algorithm to obtain new keywords using the current crawled database, and selects a main topic for each page. The topics are organized as rooted trees with all the related keywords under the main topic. Once the main topic of the clicked page is identified it is matched against the existing topics. If it matches with any of the existing topics, the new topic is merged with the most matching topic. The next phase is to incorporate the increased user priority for the topic while computing the page rank. The topics are ranked according to the user's personal preferences learned from the user's profile. For example, by tracking down that a user frequently clicked on pages describing sports topics, the system will rank all topics related to sports higher than the other topics, and recommend to the user his/her potential favorite link.

### 3.1 Online Profiling

Online activities are tracked by analyzing web history from the browser. It involves examining and analyzing what users have been viewing on the Web through the browser such as websites visited, searches conducted and web-based e-mail. Web history provides a general profile of the user, portraying the user's behavioral patterns. The history files produced by the browsers are not in a human readable format and parsing them requires external tools. For example, Internet Explorer stores usage history in several index.dat files and Firefox stores its web history in SQLite databases. Similar to Firefox, Chrome uses SQLite database and history contains a table of unique URLs visited called urls, and a table for each unique visit called visits[21]. In Chrome, it is possible to trace a user's path on the web by referring URLs and visit types [22].

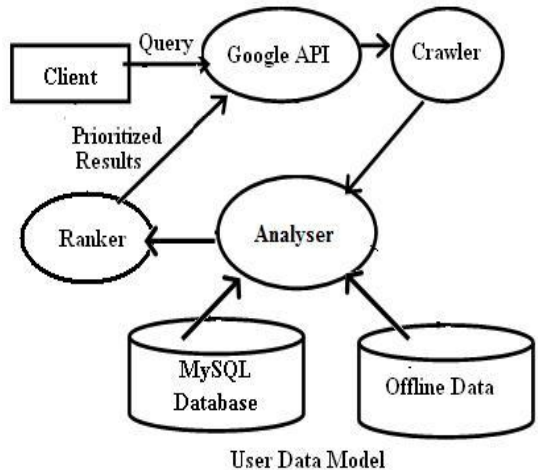

**Fig 1: The system architecture**

Google Chrome is used for this purpose since it is well cloud synchronized and is more advanced than other browsers. User's web activity is analyzed by fetching the data from database of SQLite using the default APIs of Google Chrome. Currently records of web activity such as User history, Bookmarks, Downloads, Cookies and Autosuggestion are considered for extracting the online behavior.

### 3.1.1 Window of Observation

While building the profile, the accumulation of data in the database causes two problems: Performance degradation due to oversize and difficulty to incorporate current interests of the user. Database performance degradation is especially important when SQLite is used. The performance of the database operations is inversely proportional to the size of the database. Web activities of the user are highly dependent on the temporary interests of the user. A user profile should concentrate more on the local temporary interests of the user than the global permanent interests of the user. To solve the above problems, a window of observation (WOB) is maintained. All the calculations are done on the window of observation, which is a small interval of time in the history that is used in the user modelling. Normally, data from the current WOB only is used for analysis. WOB is set according to the database size. Default limit for database size is taken as 100 MB. When the database size increases over 100 MB, a new WOB is created. During the analysis of the user interests, to incorporate the permanent global interests of the user in the profile, previous WOB with necessary weight is also considered in addition to current WOB. This will increase the performance and precision.

### 3.1.2 Capturing user behavior from search Queries

User search pattern also exhibit implicit user interests. This is captured using the search pattern extraction algorithm which is executed at frequent intervals.

Algorithm 1: Search Pattern Extraction
 a) Fetch the search queries from history database
 b) Find the frequency of the search query
 c) Rank the search query and convert it into percentile

### 3.1.3 Document Visit Profiling

Visited/Clicked pages are of primary interest to the user. Three factors from each visited document is considered by the personalization engine. Topic priority, Keyword relevance and frequency of visits to the respective URL are used in prioritizing the search results. Algorithm Grade_URL computes the prioritized rank of the visited sites.

Algorithm 2: GRADE_URL
Important Terms :
U={url1,url2,url3,……urln} U set of all visited URLs. Assume that there are n visited URLs
Frequency ={fu1,fu2,fu3…..fun} Fu set of frequencies of visits of URLs
LMTu={LMTu1,LMTu2……LMTun} LMTu is the set of last modified time of URLs
Visit_Duration={VDu1,VDu2…….VDun} Visit duration time of URLs

 a) Fetch set of URLs U from the history database
 b) Select the URLs under the current window of observation. Rank the URLs using LMTu, with higher rank assigned to the relatively new URLs. Compute Fressness_Value in a scale of 0 to 1 from the percentile score computed from the rank of the URL.
 c) Rank all the URLs U based on frequency of visit Fu. Find the percentile frequency for each URL.
 d) Rank all the URLs U based on visit duration VDu. Ranks are assigned in such a way that URLs with long visit duration are assigned high ranks. Find the percentile duration for each URL.
 e) If URL is clicked set Typed=0. If URL is typed set Typed=1
 f) Total URL Grade RUi=(Frequency + Visit_Duration + Typed)* Fressness_Value

### 3.1.4 Personalized Topic Grading

The aim of topic ranking is to rank a user's potentially interesting topics higher. Topic graph contains links to all related topics. Each topic has a weight associated with it which is proportional to the frequency of occurrence of the topic. When a topic is created its weight is 1. Whenever a new page is grabbed, the weights of all related topics will be incremented by a factor proportional to the similarity of the topic with the page. The undirected edge between the nodes X and Y represents the similarity between topic x and topic y. When the weight of a topic x is incremented by k, weights of all related topics also will also be incremented by a factor proportional to the weight of similarity edge to the related topic. Each topic will have associated set of keywords represented as a keyword tree. The keyword tree is maintained as a max heap. Root of the tree is identified as the topic. Each keyword will be assigned a relevance/rank based on its frequency of occurrence related topics are clustered by checking the similarity between the keyword trees. Topic names as represented as graphs. When a new topic is introduced, a default weight is assigned to the topic. Edges are added between topics depending on the similarity between the topics. Whenever a user clicks on a topic or related topic, weight of the topic increases. An edge is removed if the weight of the edge is less than a particular threshold value.

Three WOB's are maintained for topics: Present, Prev(for Previous) and Old. Present and Prev together forms the Current WOB. Grade of a topic is determined by considering the weight of the topic in the Present, Prev and Old.

Algorithm 3: Topic clustering
a) Each topic is identified as a node in the topic graph for Present WOB.
b) If the identified topic is present in the existing topics, merge the trees.
c) If the identified topic is not present in the existing topics it is matched against the existing keyword trees to check similarity. If the similarity is more than a particular threshold, merge the trees and select a topic name among these and increment topic value.
d) If the topic is identified as new, it should be added to the topic graph. .
    i. If one page contains a hyperlink to another page, it is represented as a weighted edge between the corresponding topic nodes. If the edge already exists, weight of the edge between the nodes is increased.
    ii. Include weighted edges between the new topic and similar topics by calculating similarity of the keyword trees. If the edge already exists, weight of the edge between the nodes is adjusted.
    iii. Topic_value = (frequency + duration)* sim_factor // sim_factor is the similarity of the URL with the topic- obtained by comparing the keyword tree with the topic
e) A fraction of the Topic_value of the current topic proportional to the edge cost is propagated to all related topics.
f) Weights of the topics in the old WOB are updated considering a freshness_factor of 0.9. Irrelevant nodes and edges are removed. old WOB and Prev WOB are merged
g) All connected components form a cluster.

### 3.1.5 Finding Topic Weights
A personalized grade is assigned to each topic. The topic value represents the prioritized grade of the topic for a particular user.
For each topic Compute Topic Value= $\alpha$*.75+ $\beta$*.25 where $\alpha$ is the Topic weight obtained from Cuurent WOB(Prev+Present) and $\beta$ is the Topic weight obtained from Old_WOB.

### 3.1.6 Finding Keyword Weights
In addition to the topic graph, keyword tree a keyword database(list) is maintained. The keyword list is used to ensure that a particular keyword has already occurred or not.

Algorithm 4: Keyword_Grading // Algorithm for extracting and assigning grades to keywords
a) Select URL Ui from U
b) Fetch the URL and extract the keywords
c) Find Frequency of keywords, Rank it based on Frequency. Calculate percentile
d) Select the keywords with percentile greater than the default value € (e.g. 70)
e) Include the keywords in the Keyword database. If the keyword is already present in the Keyword database increase the frequency by 1. Else insert the new keyword with frequency 1.

## 3.2 Offline Profiling
Offline activities are fetched by analyzing the files residing in the users system. To analyze the files, we can use windows indexing service API which is provided by Microsoft for developers. Indexing service uses filters to read and extract contents from files. Caching and use of fuzzy algorithm improves the speed of tracking the files and the folders. Contents extracted from the files are arranged in the form of catalogs for efficient retrieval.

In the current implementation, to reduce the complexities of handling large amount of data, for the offline profile, only selected files are used. The user is allowed to choose the files which are important to him. We not only take the names of files but also track the textual content within documents and text formats and also analyze the metadata in case of other formats. We then rank these documents on basis of keywords. All the data is synchronized with the database on the server. In the server an analyzer is implemented which parse the keywords and rank them accordingly. There is also a domain parser which analyses the domain names and the URL path and find top visited websites. A Graphical User Interface is also created for users to manage the user data model from user's side.

## 3.3 Personalized Web Searching
For the conventional web searching, the search engine returns the same results for different users in spite of the different interests possessed by the different users. It does not provide tailored search results. But by making use of the user model we can personalize the results for different users.

Personalization improves the search results in many ways. Incorporation of search and click behaviour and content based priority assignment makes the personalised system effective and efficient for search. As mentioned above the user data model has a set of keywords with related priorities. The obtained search result holds the most relevant links related to the search query. Along with the links a small description of the sites are also available. The links and the description are parsed to obtain the keywords. The obtained keywords are compared with the user data model and a summation of priorities is obtained. After obtaining the total priorities the relative priorities are computed. The search results are ranked as per their relative priority and the links are listed in the descending order of the rank as in Figure 2.

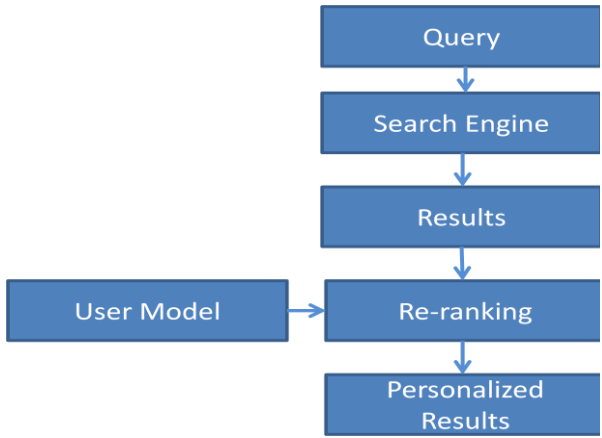

**Fig 2: The personalized search engine.**

### 3.3.1 Re-Ranking the Results

The algorithm uses search results obtained from the Google search engine and refines the obtained results by re-ranking. The user is allowed to select from a list of keyword profiles and the selected profile keywords are included in the search. An Initial list of search results, with a default value of 100 results, are collected and stored in the search bank for further processing. For each search result, the URL, topic, keywords and web rank are passed to the personalized re-ranker. Once the search results are received, the results in the search bank are re-ranked according to the personal interests of the user. The interests are represented in terms of URL visit history, frequently processed keywords and topics, stored documents (keywords and topics). Re-ranking is achieved through a matching function which calculates the degree of similarity between each search result and the user profile.

Algorithm 5: Personal_Search

a) For each search result in the search bank, perform keyword tree generation and topic name identification.

b) Compute the grade of each URL as

Grade = a* $U_G$ + b* $K_W$ + c*$T_V$ +  d* $O_V$ + e* $W_R$ + f* $S_G$
where $U_G$ - URL_Grade; $K_W$- Keyword_Weight;
$T_P$ -Topic_Value; $O_V$ - Offline_Value; $W_R$ - Web_Rank;
$S_G$ - Search_Keyword_Grade and a+b+c+d+e+f =1;  a ,b, c, d, e, f  are modified  according to the search behavior of the particular user.

c) List the results greater than a threshold value which will give personalized search result(ranking) in the descending order of scores obtained.

## 4.  PERFORMANCE EVALUATION

Usually  the  search  engine  performance  is  measured  in  terms  of precision  and  recall[23][24][25].  Precision  is  the  fraction  of retrieved instances that are relevant, while recall is the fraction of relevant instances that are available. Both give an idea about the relevance of search results. But here the personalized engine just refines the results returned by the Google search engine. So the precision and recall curves are same as that of the typical respective search engine, considering the fact that a result will appear in the personalized search result only if it is present in the original search result. Personalization only improves the relative ranking of the relevant documents.

Initial  experiments  to  evaluate  the  personalisation  engine  were conducted  using  the  user  model  of  two  computer  science teachers.  Though  both  of  them  are  computer  science  teachers, their subject interests were diverse. Result of the experiments is shown  in  Figure  3.  The  graph  depicts  the  original  search  rank (before  personalisation)  of  the  potential  pages  of  each  user.  X axis shows the revised page rank after re-ranking and Y axis shows  the  respective  unrevised  rank.  From  the  user  click  pattern, it  was  observed  that  the  utility  of  the  first  page  and  user satisfaction has improved much when personalization is applied.

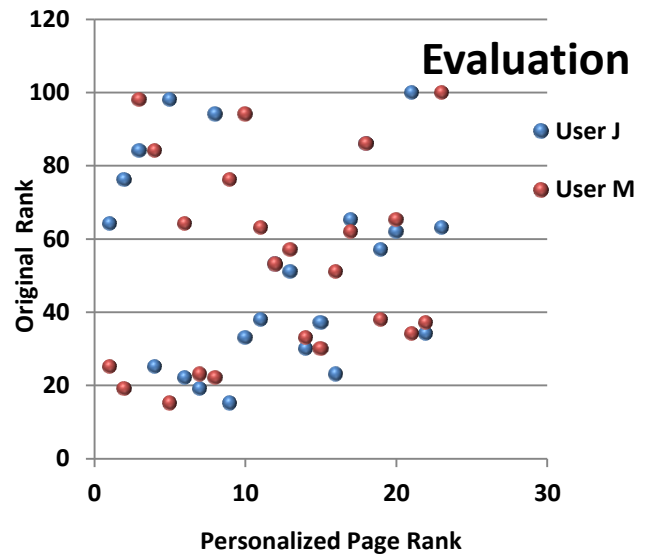

**Fig 3: Effect of personalization for potential pages.**

## 5.  CONCLUSION

Personalization improves accuracy and utility of retrieved results and user satisfaction. When a search was performed using the keyword "journal", the first relevant result for a computer professional is the "Linux Journal" with rank 13 as on January 2013. For the search string "database journal", only 2 of 10 in the first page are found to be useful for computer science personnel. But when personalized engine is used, it was found that most relevant results were brought to the initial positions with marginal improvement in utility. It is clear that the personalized system helps people to find what they are looking for easily without wasting time on unwanted sites. Contributions of the paper are: Development of the algorithm for extracting relevant keywords, Incorporation of the concept WOB for speed improvement and experimental evaluation for the impact of personalization. The best application of the system comes in the field of advertising. With this system, advertisements that a person sees can be personalized according to his interests with considerations of the seasonal variations in interest. Thus advertisers can get good returns and publishers can get high click through rates. Mails can be prioritized according to the user's personality. Music search, app search, feed reader, newspaper search are other areas where personalization can be applied.